\documentclass{elsart}
\usepackage[dvips]{color}
\usepackage{textcomp}
\usepackage{amsmath}
\usepackage{amssymb}
\usepackage{graphicx}
\usepackage{wrapfig}
\usepackage[ps2pdf,a4paper=false,letterpaper=true,colorlinks=true,urlcolor=blue,citecolor=magenta,bookmarks=false]{hyperref}
\usepackage{epsfig}
\input epsf

\begin{document}
\def\pitildem{$\tilde{\pi}$\ }
\def\pitilde{\tilde{\pi}}
\def\lambdacm{$\lambdac$\ }
\def\lambdac{\Lambda_c^+}
\def\lambdacbm{${\overline \Lambda}{}_c^-$\ }
\def\lambdacb{{\overline \Lambda}{}_c^-}
\newcommand{\alm}{$\alpha_{\Lambda_c}$\ }
\newcommand{\albm}{$\alpha_{\overline{\Lambda_c}}$\ }
\newcommand{\al}{\alpha_{\Lambda_c} }
\newcommand{\alb}{\alpha_{\overline{\Lambda_c}} }
\newcommand{\alambda}{\alpha_{\Lambda} }
\def\costm{$\cos\theta$\ }
\def\cost{\cos\theta}
\def\signal{\Lambda_c^+\to \Lambda\pi^+}
\def\signalm{$\Lambda_c^+\to \Lambda\pi^+$\ }
\def\signalb{\overline {\Lambda}{}_c^-\to {\overline \Lambda}\pi^-}
\def\signalbm{$\overline {\Lambda}{}_c^{-}\to {\overline \Lambda}\pi^-$\ }
\def\reflem{$\Lambda_c^+\to \Sigma^0\pi^+$\ }
\def\reflebm{$\overline{\Lambda}{}_c^- \to \overline{\Sigma}^0 \pi^-$\ }
\def\alrefle{\alpha_{\Lambda_c\to \Sigma^0\pi}}
\def\alreflem{$\alpha_{\Lambda_c\to \Sigma^0\pi}$\ }
\def\A{$\mathcal{A}$\ }
\def\valnal{-0.73}
\def\ervalnal{0.21}
\def\valnbaral{0.84}
\def\ervalnbaral{0.22}
\newcommand{\valal}{-0.79}
\newcommand{\erstvalal}{0.15}
\newcommand{\ersyvalal}{0.13}
\newcommand{\albias}{-0.01}
\newcommand{\valalbis}{-0.78}
\newcommand{\erstvalalbis}{0.16}
\newcommand{\valalnofix}{-0.84}
\newcommand{\ervalalnofix}{0.14}
\def\Avalnofix{0.06}
\def\erAvalnofix{0.17}
\def\Aval{-0.07}
\def\erstAval{0.19}
\def\ersyAval{0.12}
\def\Abias{0}
\def\Avalbis{-0.07}
\def\erstAvalbis{0.19}
\def\BBR{\mathrm{BR}_{\Sigma^0}}
\def\BBRm{$\mathrm{BR}_{\Sigma^0}$}
\def\BR{0.98}
\def\erBR{0.11}
\def\BRn{1.14}
\def\erBRn{ 0.15}
\def\BRnbar{0.81}
\def\erBRnbar{0.16}
\begin{frontmatter}

\title{Study of the  decay asymmetry
parameter and \emph{CP} violation parameter in
the \signalm decay.}

The FOCUS Collaboration\footnote{See \textrm{http://www-focus.fnal.gov/authors.html} for
additional author information.}

%\collaboration{The FOCUS Collaboration
%\footnote{See \textrm{http://www-focus.fnal.gov/authors.html} for 
%additional author information.}
%}

\author[ucd]{J.~M.~Link}
\author[ucd]{P.~M.~Yager}
\author[cbpf]{J.~C.~Anjos}
\author[cbpf]{I.~Bediaga}
\author[cbpf]{C.~Castromonte}
\author[cbpf]{A.~A.~Machado}
\author[cbpf]{J.~Magnin}
\author[cbpf]{A.~Massafferri}
\author[cbpf]{J.~M.~de~Miranda}
\author[cbpf]{I.~M.~Pepe}
% ...

\address[ucd]{University of California, Davis, CA 95616}
\address[cbpf]{Centro Brasileiro de Pesquisas F\'\i sicas, Rio de Janeiro, RJ, Brazil}
% ...
\address[fnal]{Fermi National Accelerator Laboratory, Batavia, IL 60510}
% ...

%{\Large FOCUS AUTHOR LIST}
%\author{Author List}
\begin{abstract}
Using data from the FOCUS (E831) experiment at Fermilab, we present a new 
measurement of the weak
decay-asymmetry parameter \alm in  \signalm decay.
Comparing particle with antiparticle decays, we obtain the first measurement
of the \emph{CP} violation parameter
$ \mathcal{A} \equiv { \alpha_{\Lambda_c}+\alpha_{ \overline{\Lambda_c} }
 \over \alpha_{\Lambda_c}-\alpha_{\overline{\Lambda_c}} }$.
We obtain $\al = \valalbis\pm \erstvalalbis\pm \ersyvalal$ and
 $\mathcal{A} = \Avalbis \pm \erstAvalbis \pm \ersyAval $ where
 errors are  statistical and  systematic.
\end{abstract}
\end{frontmatter}

\section{Introduction}

Parity violation in the weak decay of a spin $1\over 2$ hyperon into
a spin $1\over 2$ baryon and a pseudoscalar meson is well known.
An example is $\Lambda \to p \pi^-$~\cite{kallen}, where the angular
distribution of the proton in the $\Lambda$ rest frame is given by
\begin{equation}
{dW\over d\cos\theta} = {1\over 2} (1+ P \alpha _{\Lambda} \cos\theta)
\label{uno}
\end{equation}
where {\it P}
is the polarization of  $\Lambda$ along the $z$ direction, $\theta$ is
the polar angle and $\alpha_{\Lambda}$ is the weak decay-asymmetry parameter.
The latter is defined as
\begin{equation}
\alpha_{\Lambda} \equiv { 2 {\mathcal Re } (A_+^*  A_-) \over |A_+|^2 + |A_-|^2 }
\label{weakampli}
\end{equation}
where $A_+$ and $A_-$ are the parity-even and parity-odd decay
amplitudes. In a
non-relativistic picture, they correspond to the $L=0$ and $L=1$
orbital angular momenta of the proton-$\pi$ system
respectively.

Eqn. (\ref{uno}) can be generalized for
 the case of a baryon double decay chain like
\signalm with $\Lambda \to p \pi^-$ in which each
baryon weak decay is a ${1 \over 2 }^+ \to {1 \over 2 }^+ + {0^- }$
process. If the \lambdacm
is produced unpolarized, Eqn. (\ref{uno}) for the $\Lambda$\ decay becomes
\begin{equation}
{dW\over d\cos\theta} = {1\over 2} (1+ \alpha_{\Lambda_c}\alpha_{\Lambda}\cos\theta)
\label{due}
\end{equation}
where $\alpha_{\Lambda_c}$ is the weak-decay asymmetry parameter of the
\signalm process and $\theta$ is the helicity
angle of the proton in the $\Lambda$ rest frame, {\it i.e.}
it is the supplement   of the
angle  ({\it i.e.} $\pi$ minus the angle)
in the $\Lambda$ rest frame between the proton and the
direction of the $\lambdac$.
% I think the language is clear and that figure can be
% omitted. (see fig.~\ref{helicity}).
  For this
decay of the $\lambdac$, this is the same as the angle, in the $\Lambda$\ rest
frame, between the proton and the $\pi^+$ decay product of the
$\lambdac$, and the supplement of this angle is used in this analysis.
The distribution of the helicity angle, $\theta$, determines the product
$\alpha_{\Lambda_c}\alpha_{\Lambda}$, and,
since $\alpha_{\Lambda}$ is
known, (in this analysis we assume
$\alpha_{\Lambda}=-\alpha_{\overline\Lambda}=0.642\pm 0.013$~\cite{PDG}),
one can extract $\alpha_{\Lambda_c}$.

If \emph{CP} were conserved exactly, $\alpha_{\Lambda_c}$ of the
\signalm process  would be the negative
of $\alpha_{\overline{\Lambda_c}}$ in
\signalbm decay.  Just
as with the $\Lambda$ decay, there is the possibility of \emph{CP} violation in 
\lambdacm decay (a weak decay
with possible final state interactions).  
In this paper we 1) measure $\alpha_{\Lambda_c}$\ and
$\alpha_{\overline{\Lambda_c}}$\ separately, 2) use them for the first measurement
of the \emph{CP} asymmetry parameter
\begin{equation}
\mathcal{A} \equiv { \alpha_{\Lambda_c}+\alpha_{ \overline{\Lambda_c} }
 \over \alpha_{\Lambda_c}-\alpha_{\overline{\Lambda_c}} }
\label{cp_viola}
\end{equation}
and 3) having established that the difference is negligible within our errors,
combine the data for particle and
antiparticle decays to obtain
the best value of the asymmetry parameter.  In this analysis
the \lambdacm is assumed to be unpolarized
 and thus the $\Lambda$\ longitudinal
polarization (polarization along the $\Lambda$ momentum direction
in the \lambdacm rest frame)
is $\alpha_{\Lambda_c}$.
As discussed later, if the polarization is the
maximum allowed in our data, the effect on our measurement is negligible.

FOCUS is a charm photoproduction experiment, 
an upgraded version of E687~\cite{spectro},  which collected 
data during the 1996--97 fixed target run at Fermilab.
 Electron and positron beams
 obtained from the
$800~\mathrm{GeV/c}$ 
Tevatron proton beam produce, by means of bremsstrahlung, a photon beam
 (with
typically $300~\mathrm{GeV}$ endpoint energy)
 which
interacts with a segmented BeO target~\cite{photon}. The mean photon energy for triggered
events is $\sim 180~\mathrm{GeV}$. A system of three multicell threshold \v{C}erenkov
counters
is used to perform
 the charged particle identification, separating kaons from
pions up to $60~\mathrm{GeV}/c$ of momentum. Two systems of silicon microvertex
detectors are used to track particles: the first system consists of 4 planes
of microstrips interleaved with the experimental
target~\cite{spectro,WJohns}
 and the
second system consists of 12 planes of microstrips located downstream of the
target. These detectors provide high resolution in the transverse plane
(approximately $9~\mu\mathrm{m}$ on the track position),
 allowing the identification and
 separation of charm
production  and decay vertices. The charged particle
momentum is determined by measuring the deflections in two magnets of
opposite polarity through five stations of multiwire proportional chambers.

%%%%%%%%%%%%%%%%%%%%%%%%%%%%%%%%%%%%%%%%%%%%%%%%%%%%%%%%%%%%%%%%%%%%%%%%%%%%%%%%%%%
%

\section{Analysis of the decay mode \signalm +
charge conjugate.} %
%%%%%%%%%%%%%%%%%%%%%%%%%%%%%%%%%%%%%%%%%%%%%%%%%%%%%%%%%%%%%%%%%%%%%%%%%%%%%%%%%%%
Unless explicitly stated otherwise, both the particle and its charge 
conjugate
are implied.
The selection of \lambdacm events begins with the identification of
$\Lambda$\ candidates on the basis of vertexing
and loose \^Cerenkov cuts on the pion and proton decay particles as described in
detail elsewhere~\cite{nimvee}.
The tracks of these charged daughters are used to form the decay
vertex of the $\Lambda$ and to determine its flight direction and
momentum.  The momentum of the
$\Lambda$ is used together with the momentum of a $\pi^+$
(or with the momentum of a $\pi^-$ in the case of a $\overline \Lambda$)
 to find the  \lambdacm (\lambdacbm) vertex and its
momentum.  We call this pion $\pitilde$.
The $\Lambda$'s used in this analysis can be divided into
two categories: the $\sim 10$\%
which decayed before the microstrip detector
and the large majority that decayed
after the first microstrip plane.
For the former,  both the $\Lambda$  decay vertex
and the momentum vector are measured precisely by
the FOCUS spectrometer; for the latter only the momentum vector 
is measured precisely.
Thus different algorithms are used
to
find the \lambdacm decay vertex and the production vertex
in these interactions.
For the  $\Lambda$'s decaying upstream of the
the microstrips, both the vertex position and the direction of the
$\Lambda$ are used in combination with a fully reconstructed \pitildem 
track.
The confidence level that the \pitildem and the $\Lambda$
originated from the same \lambdacm
decay vertex is computed and required
to be $> 1$\%.
The \lambdacm candidate decay vertex and momentum are calculated and used as
the initial track for
a \textit{candidate driven vertex
algorithm}~\cite{spectro}.
 The momentum and position of the resultant \lambdacm candidate are used as
a \textit{seed} track to intersect the other reconstructed tracks and to 
search for a production vertex. The confidence level of this vertex is
required to be greater than 1\%.

For the $\Lambda$'s decaying downstream of the first plane of the microstrip,
we use the momentum information from the $\Lambda$ decay and the silicon track
of \pitildem
to form a candidate \lambdacm momentum vector.
 This vector and the \pitildem track form a plane in which the momentum vector
of the candidate \lambdacm lies. The transverse distance from this plane
is calculated for candidate production vertices which are formed by at least two other
silicon tracks and the confidence level that the vertex lies in
the plane is computed.
The \lambdacm trajectory is forced to originate at this production vertex
and the confidence level that it verticizes with
\pitildem is calculated.
If this confidence level is greater than 2\%, this vertex is taken as the
decay vertex of the \lambdacm and used with the \lambdacm momentum
as the  \textit{seed} track for a candidate driven search for the production
vertex.  The confidence level of this
vertex is required to be greater than 1\%.

For both types of $\Lambda$'s, once the production  vertex is 
determined, the distance $l$ between the production and decay
(\lambdacm decay) vertices and its error $\sigma _{l}$ are computed. 
The quantity $l/\sigma _{l}$ is an unbiased measure of the 
significance of 
detachment between the production and decay vertices. This is
the most important variable
for separating charm events from non-charm prompt backgrounds.
In this analysis a cut of $l/\sigma _{l} > 3$ has been imposed on the
selected events.
Signal quality is further
enhanced by imposing tighter \v{C}erenkov identification cuts on the proton
decay product of the $\Lambda$
(to eliminate
the contamination of $K_S^0$ decaying into two pions)
and $\pitilde$.
The \v{C}erenkov identification cuts used in
FOCUS are based on likelihood ratios between the various particle
identification hypotheses. These likelihoods are computed for a given track
from the observed firing response (on or off) of all the cells that are
within the track's ($\beta =1$) \v{C}erenkov cone for each of the three 
\v{C}erenkov counters. The product of all firing probabilities for all the cells
within the three \v{C}erenkov cones produces a $\chi ^{2}$-like variable 
$W_{i}=-2\ln (\mathrm{likelihood})$ where $i$ ranges over the electron, pion,
kaon and proton hypotheses~\cite{cerenkov}.
The proton track is required
to have %$\Delta _{p,\pi} \equiv
$W_{\pi }- W_{p}$  greater than 4 and %$\Delta _{p,K} \equiv 
$W_{K }- W_{p}$ greater than 0, 
whereas
the \pitildem track is required
to be separated by less than 6 units from the best
hypothesis, that is % $picon\equiv W_{min}-W_{\pi }$ (pion consistency).
pion consistency, $W_{\mathrm{min}}-W_{\pi } < 6$.
Signal quality is further
enhanced by requiring that the momentum of the \lambdacm be greater
than 40 $\mathrm{GeV}/c$, the position of the production vertex to lie within
the target fiducial volume, the proper lifetime of the candidate \lambdacm
to be less than 5 times the nominal \lambdacm lifetime, the transverse
momentum of the \lambdacm  with respect to the average beam direction be
greater than 0.2 $\mathrm{GeV}/c$ and finally by requiring that
$|\cos\omega|<0.8$, where $\omega$ is the angle between the
\lambdacm flight direction and the \pitildem flight direction
evaluated in the \lambdacm rest frame.

Using the set of selection cuts described, we obtain the
invariant mass distribution 
for $\Lambda\pi^-$ shown in Fig.~\ref{mass} with $776\pm55$ events in
the particle channel and $637\pm 34$ events in the antiparticle
channel.  The events in these plots are further divided into four slices of the
cosine of the $\Lambda$ decay angle, and the numbers of  $\Lambda\pi$
decays of the  \lambdacm above background in each of the four slices
constitute the histogram of events {\it vs.}\ $\Lambda$\ decay angle.
In order to do the fits for the numbers of decays, three additional features
must be recognized:
1) the change in slope at 2.15  $\mathrm{GeV}/c^2$, 2) the \reflem
reflection just below the  \lambdacm peak, and 3) the smooth background.

The decay mode
$\Lambda_c^+ \to \Lambda \pi^+\pi^0$ where the $\pi^0$ is not
detected contributes to the rise below 2.15 $\mathrm{GeV}/c^2$.
Examination of Monte Carlo
generated $\Lambda_c^+ \to \Lambda \pi^+\pi^0$ events verified that this
reflection becomes negligible above 2.15 $\mathrm{GeV}/c^2$ and motivates the
choice of 2.15 $\mathrm{GeV}/c^2$\ as the lower limit of the mass region
used for this analysis.

The bump from 2.15 to 2.25 $\mathrm{GeV}/c^2$ is the decay
\reflem where the $\gamma$\ from  $\Sigma^0 \to \Lambda \gamma$\ is
not seen.  Including the
$\Sigma^0$\ region provides a check on the background parameterization but requires
that the angular distributions of the $\Sigma^0$\ mode must be used in the
mass and efficiency calculations.
Because of the electromagnetic decay~\cite{gatto,primakov}\ in this chain,
the $\Sigma^0$\ decay is isotropic and, integrated over all $\Sigma^0$\ decay
angles,
the helicity angle distribution of the proton from the $\Lambda$ is flat
regardless of the initial polarization of the \lambdacm and the
value of $\alpha_{\Lambda_c \to \Sigma^0\pi}$ (the weak decay-asymmetry
parameter of the \reflem decay).  However, in this analysis, the acceptance,
efficiency, and calculation of
%and assignment to slices in
the cosine of the $\Lambda $ decay angle  from experimental data
may depend upon
the $\Sigma^0$\ decay angles.  Moreover, the helicity angle is
miscalculated as the
angle, in the $\Lambda$ rest frame, between the proton and the negative of the
direction of the pion $\pitilde$ rather than the
negative of the direction of the {\it photon}.  Thus the
distribution of observed proton helicity angles in this mass region 
could be distributed unevenly among
the \costm slices with the distribution depending on
$\alpha_{\Lambda_c \to \Sigma^0\pi}$. These effects have been included in
this analysis 
by generating Monte Carlo \reflem events and reconstructing
them as \signalm events.  The events are  generated assuming the
\lambdacm is produced unpolarized and for a set of values
of ${\rm \alpha_{\Lambda_c \to \Sigma^0\pi}}$ %=-0.94,\ and\ 0.94}$.
which covers the physical range.
The differences among the four regions of $\Lambda $\ decay angle are
small but have been included in the analysis.
%  The shapes for extreme values
% of ${\rm \alpha_{\Lambda_c \to \Sigma^0\pi}}$ are shown in Fig.~\ref{mcsigman} for
% the particle and in Fig.~\ref{mcsigmanbar} for the antiparticle decay channel.

The background shape is obtained from the wrong sign $\Lambda \pi^-$
({\it i.e.}\ the pion from the decay vertex and the nucleon from the
$\Lambda$ have the opposite charge) mass
distribution.  This mass distribution is fitted with a second degree polynomial
and the polynomial is scaled by a multiplicative factor in the subsequent fits.
The cuts used for this background distribution are the same as for the signal
except for the additional cut
$\cos\theta < 0.7$ to remove events in which
the $\pi^-$ ($\pi^+$) track associated by the tracker with a
proton to form a $\Lambda$ ($\overline{\Lambda}$) vertex
was actually a random track, and the true $\pi^-$
($\pi^+$)  track was actually
misidentified as $\pitilde$.
No signal is expected or observed
in the \lambdacm mass region.
% as shown in Fig.~\ref{wrongsign}
% and \ref{wrongsignbar}.
%
%
For the \signalm signal, two Gaussians with common
mean and different resolutions are used.
These resolutions and relative amplitudes have been obtained with
a Monte Carlo simulation
and normalized so that the sum has unit area
for each $\cos\theta$ slice. 
In this way the Monte Carlo shape of the signal is preserved, and the
mass fit varies only an overall multiplicative factor.  
\section{Fitting procedure and extraction of $\alpha_{\Lambda_c}$ and \A}
\begin{figure}[!!t]
\begin{center}
\epsfysize=6.5cm
\epsfxsize=6.5 cm
\epsfbox{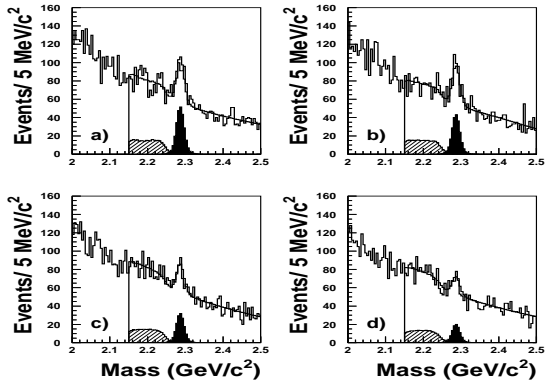}
%\vspace{0.5cm}
\end{center}
\caption{$\Lambda\pi^+$ effective mass for the \lambdacm subsample
in each \costm bin:
a) $-1<\cos\theta<-0.5$;
b) $-0.5<\cos\theta<0$;
c) $0<\cos\theta<0.5$;
d) $0.5<\cos\theta<1$.
The results of the analysis fit (solid line) are shown superimposed
on the plots.
The \signalm
 contribution to the fit is shown by the solid
histogram, the
\reflem
 contribution by the hatched histogram.
}
\label{cost_n}
\end{figure}

\begin{figure}[!!t]
\begin{center}
\epsfysize=7.cm
\epsfxsize=7.cm
\epsfbox{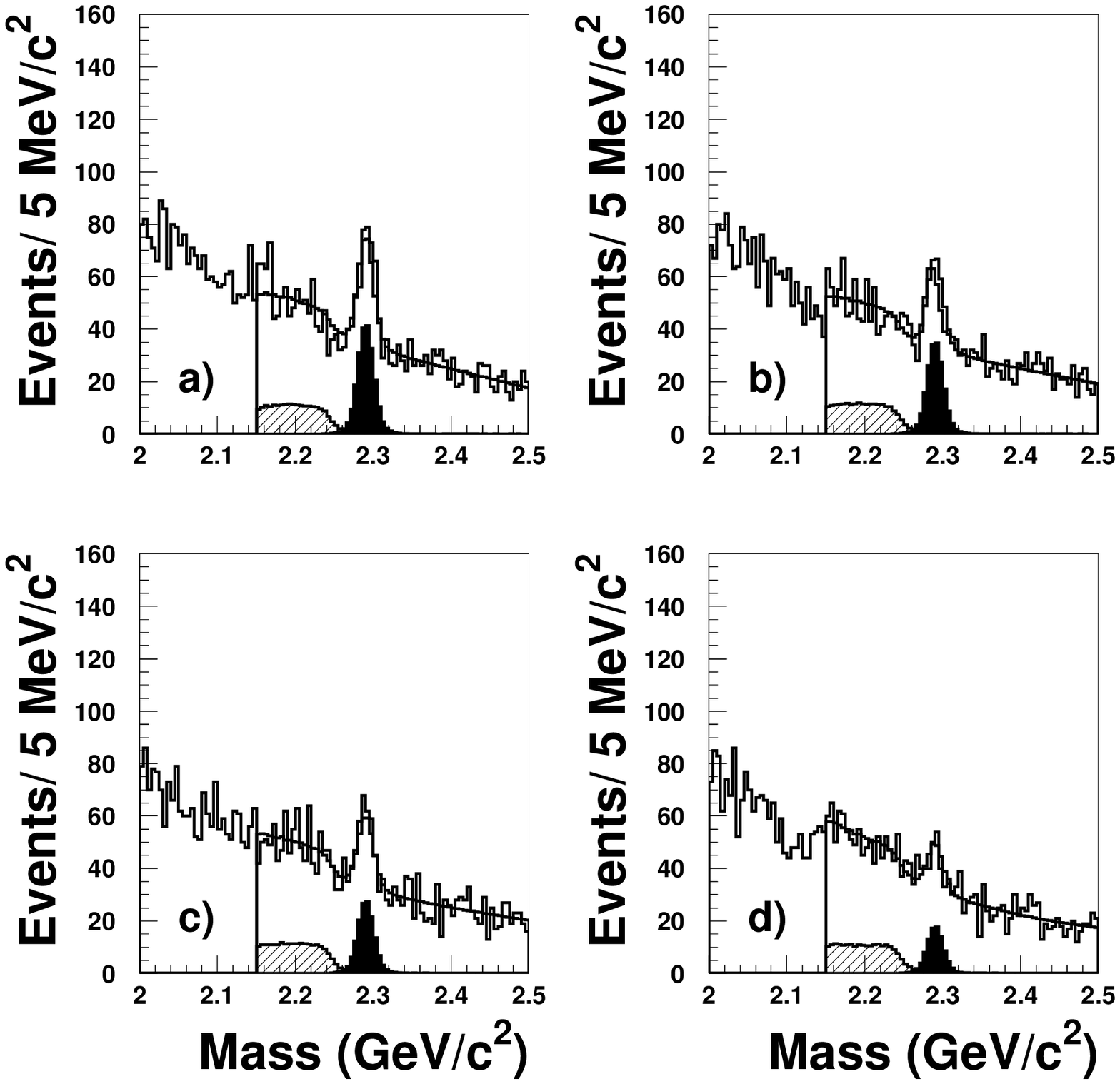}
%\vspace{0.5cm}
\end{center}
\caption{$\overline{\Lambda}\pi^-$
 effective mass for the \lambdacbm subsample
in each \costm bin:
a) $-1<\cos\theta<-0.5$;
b) $-0.5<\cos\theta<0$;
c) $0<\cos\theta<0.5$;
d) $0.5<\cos\theta<1$.
The results of the analysis fit (solid line) are shown superimposed
on the plots.
The \signalbm
 contribution to the fit is shown by the solid
histogram, the
\reflebm
contribution by the hatched histogram.
}
\label{cost_nbar}
\end{figure}

In order to extract $\alpha_{\Lambda_c}$ and \A,
the data sample is divided into particle and antiparticle subsamples and
each subsample is further divided into four equal slices of
${\cos \theta}$, spanning the range from $-1$ to +1. The $\Lambda\pi$ mass plots
of the
resulting eight subsamples
are shown in Figs.~\ref{cost_n}  and  \ref{cost_nbar}.
%  This is fig 5 and 6 as of Jun 30
The number of \signalm events above background in these plots is the
basic data needed to measure the slope and thus the product
$\alpha_\Lambda \alpha_{\Lambda_c\to\Lambda^0\pi}$\ which is called $\alpha$\ in
this discussion of the fit procedure.
These numbers  are predicted by Eqn. (\ref{predetti})
below in a
simultaneous fit to all four slices.  More specifically, if $n_i$\ is the
integral of the Gaussians signal term for the  i$^{\mathrm{th}}$ slice, then:
\begin{equation}
n_i = N_{\mathrm{sig}}\epsilon ( \theta_i,\alpha ) \int _{ \cos\theta_{\mathrm{low},i}}
^{\cos\theta_{\mathrm{up},i}}
{1\over 2}(1+\alpha\cos\theta)d\cos\theta
\label{predetti}
\end{equation}
where $\cos\theta_{\mathrm{up},i}$ and $\cos\theta_{\mathrm{low},i}$ 
are the upper and lower limits of the i$^{\mathrm{th}}$ $\cos\theta$
slice, $\alpha \equiv \al \alpha_{\Lambda}$ is a fit parameter,
$\epsilon(\theta_i,\alpha)$ is the acceptance-efficiency factor calculated with
a Monte Carlo  for the i$^{\mathrm{th}}$ \costm slice and  $N_{\mathrm{sig}}$ is the
total number of \signalm events produced in the FOCUS experiment (before
correcting for
apparatus acceptance).  Since $\epsilon$ depends upon $\alpha$, it has been
calculated with Monte Carlo for a set of values of $\alpha$ and interpolation
provides the continuous function needed for the fit.

The amplitude of the  $\Sigma^0$ reflection is parameterized as the
branching ratio:
\begin{equation}
\BBR \equiv
 {\Gamma(\Lambda_c\to \Sigma^0\pi) \over \Gamma(\Lambda_c\to \Lambda\pi)}
 = {N_{\mathrm{refl}} \over N_{\mathrm{sig}}}
\end{equation}
and $N_{\mathrm{refl}}$ multiplies the Monte Carlo shape of the reflection.  The Monte Carlo
shape includes the efficiency and is normalized to one event before acceptance
corrections.  Since the acceptance corrections depend slightly on  
$\alpha_{\Lambda_c \to \Sigma^0\pi}$, a table is constructed and linear
interpolation is used to permit calculations with different asymmetry parameters.

The background is parameterized by a single parameter $K_i$ for each slice which
multiplies the wrong-sign background.

In summary, the free fit parameters are~: $\alpha$, \alreflem,
$N_{\Lambda_c}\equiv \sum_{i=1}^4 n_i$, \BBRm\ and the four $K_i$'s.
Finally, the fit is performed maximizing the likelihood function
\begin{equation}
\mathcal{L} = \prod_{i=1}^4 \prod_{j=1}^{\mathrm{mass\ bins}}e^{-\mu_{j,i}}
\frac{(\mu_{j,i})^{s_{j,i}}}{(s_{j,i})!}
\end{equation}
where the sum of the three terms just described is $\mu_{j,i}\equiv$ predicted
events in the j$^{\mathrm{th}}$ mass bin of the
i$^{\mathrm{th}}$ mass plot ($i=1,...,4$, corresponding
to each \costm slice) and $s_{j,i}\equiv$ the number of experimental events
falling in the j$^{\mathrm{th}}$ mass bin of the
i$^{\mathrm{th}}$ mass plot.

\begin{figure}[!!t]
\begin{center}
\epsfysize=7.5cm
\epsfxsize=7.5cm
\epsfbox{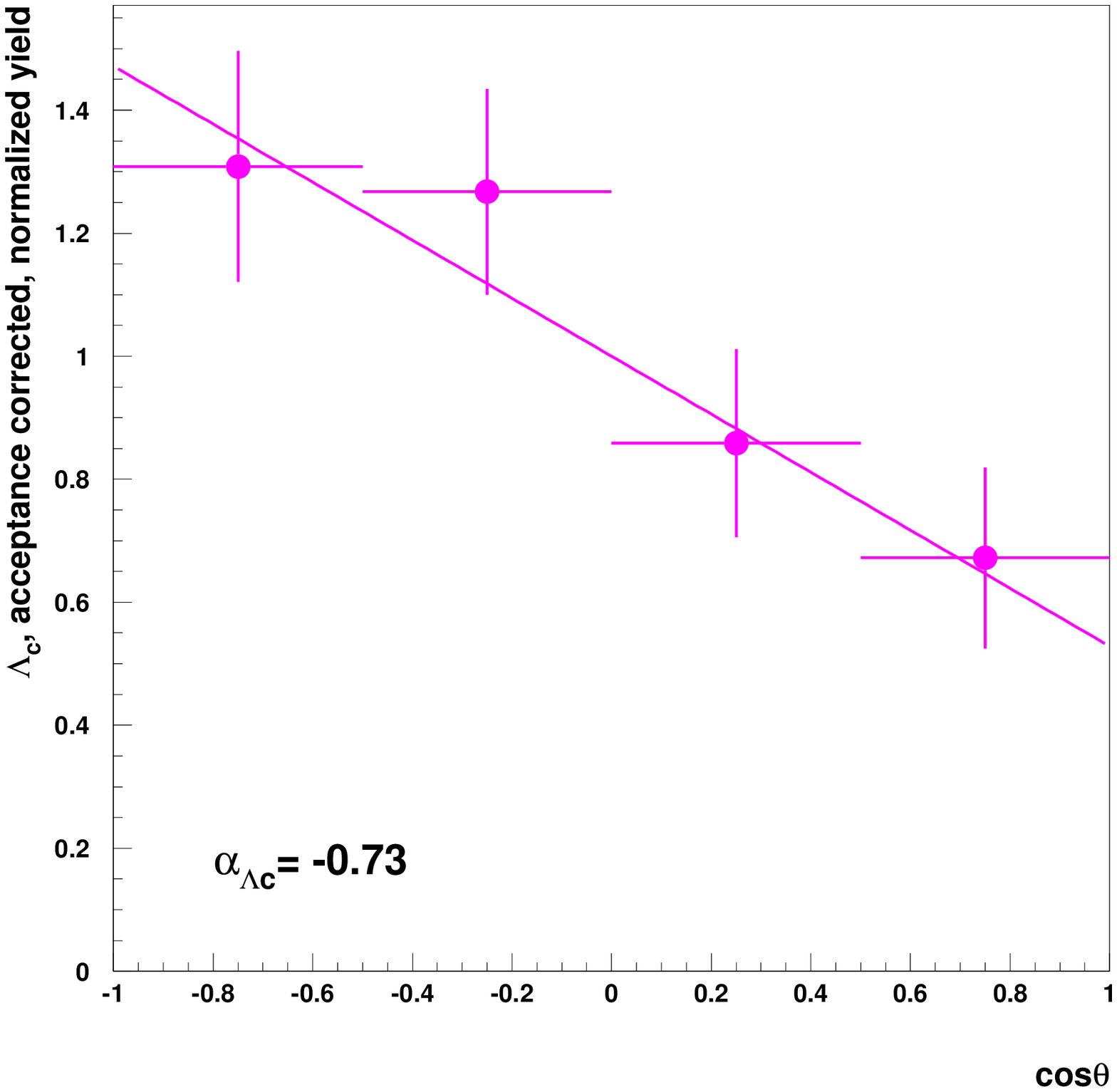}
%\vspace{0.5cm}
\end{center}
\caption{
Visualization of the fit result for the \signalm sample.
The way the values of the solid circles
and their errors are obtained are explained in the text.
The solid straight line is $1 + \al \alpha_\Lambda \cost_i$
where $\cost_i$ is the value in the middle of the i$^{\mathrm{th}}$ bin
and $\al\times\alpha_\Lambda$ is the value returned by the fit.
}
\label{plotmedn}
\end{figure}

\begin{figure}[!!t]
\begin{center}
\epsfysize=7.5cm
\epsfxsize=7.5cm
\epsfbox{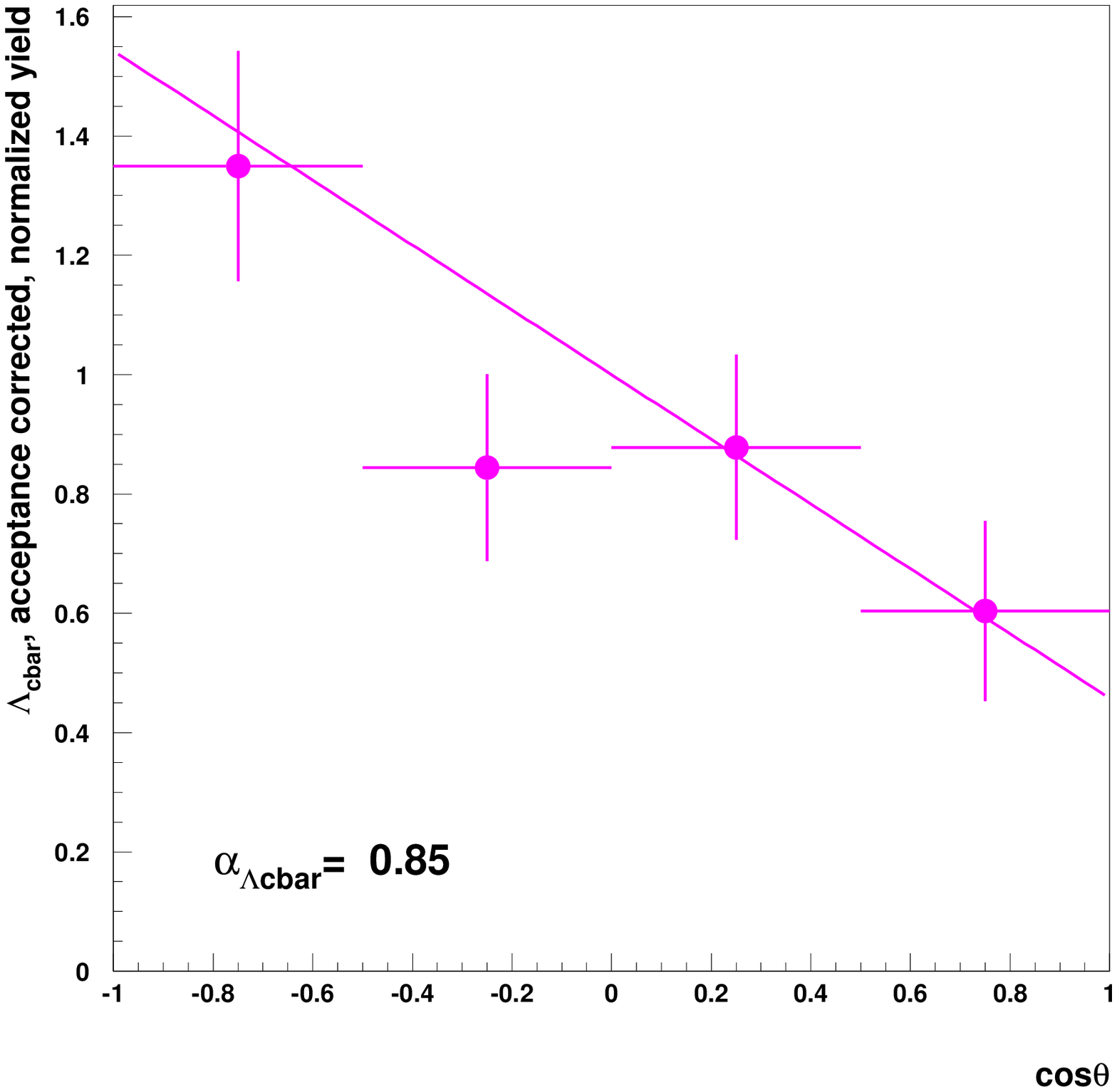}
%\vspace{0.5cm}
\end{center}
\caption{
Visualization of the fit result for the \signalbm sample.
The way the values of solid circles
and their errors are obtained are explained in the text.
The solid straight line is $1 + \alb \alpha_{\overline \Lambda} \cost_i$
where $\cost_i$ is the value in the middle of the i$^{\mathrm{th}}$ bin
and $\alb \alpha_{\overline \Lambda}$ is the value returned by the fit.
}
\label{plotmednbar}
\end{figure}

\noindent
The fit results are:
$\al = \valnal\pm \ervalnal$ (with a $\chi^2/d.o.f.$ of 1.1) and
$\alb = \valnbaral \pm \ervalnbaral$ (with a $\chi^2/d.o.f.$ of 1.0)
where the errors are statistical. These numbers lead to
$\mathcal{A} = \Aval\pm \erstAval$. Assuming \emph{CP} invariance in this
decay and taking a weighted average (after changing the
sign of $\alb$), one obtains
$\al = \valal \pm \erstvalal$.
The relative branching ratios are
$\BBR = \BRn \pm \erBRn$ for particles and
$\BBR = \BRnbar \pm \erBRnbar$ for antiparticles, leading
to a weighted average of
$\BBR = \BR \pm \erBR$
The \alreflem determined by the fit were at the edge of the physical range
both for the \lambdacm ($\alrefle=-1$) and for
 the \lambdacbm ($\alpha_{\overline{\Lambda}_c\to \overline{\Sigma}^0\pi}$)
sample,
and with a very large error of 50\%.
In Figs.~\ref{cost_n} and \ref{cost_nbar} the results of the fit are shown
for the particle and antiparticle sample respectively, with the contribution
of the \lambdacm signal and the \reflem identified. The amplitudes of the
signal and of the reflection have been computed from parameters of the global
fit over all four slices.  A further visualization of the quality of fits
are shown in
Figs.~\ref{plotmedn} and \ref{plotmednbar}.  The number of signal events 
in
each slice has been computed by subtracting the background and reflection from the
data histograms in the region between 2.255 and 2.315 $\mathrm{GeV}/c^2$.
The amplitudes of the background and reflection were taken from the global fit.
The resulting numbers are then corrected for acceptance and normalized
for comparison with the function
\begin{eqnarray}
y_i = 1+\al \alambda \cost_i
\label{formula}
\end{eqnarray}
where $y_i$ are the subtracted and corrected data points in the 
i$^\textrm{th}$ \costm bin and
$\cost_i$ is the value of \costm in the middle of the i$^\textrm{th}$ bin.
The solid line is the function in (\ref{formula}) with the value of
$\al \alambda$ returned by the fit.
%
%%%%%%%%%%%%%%%%%%%%%%%%%%%%%%%%%%%%%%%%%%%%%%%%%%%%%%%%%%%%%%%%%%%%%%%%%%%%%%%%
%
%                                  systematic checks
%
%%%%%%%%%%%%%%%%%%%%%%%%%%%%%%%%%%%%%%%%%%%%%%%%%%%%%%%%%%%%%%%%%%%%%%%%%%%%%%%%
\section{Systematic checks}
Many checks have been performed in order to assess the systematic errors
on \alm and \A.

In this analysis we assume the \lambdacm is produced unpolarized.
We think this is a safe assumption for the FOCUS experiment. In fact
if the \lambdacm\ were polarized the 
$\Lambda$\ polarization would be larger than $\alpha_{\Lambda_c}$\ in one
hemisphere about the \lambdacm\ polarization and smaller in the other hemisphere,
 but the
 polarization averaged over all \lambdacm decay angles remains
$\alpha_{\Lambda_c}$.
Even a large, asymmetric error in the acceptance correction and a
polarization
as large as 0.5 would contribute negligibly to the systematic error.
%
%
%average polarization remains very close to  $-\alpha_{\Lambda_c}$.  Specifically,
%in the extreme case of
%\lambdacm polarization of 0.4 and $\alpha_{\Lambda_c}=-0.6$, the average
% polarization
%differs from $\alpha_{\Lambda_c}$\ by only 0.02, very much less than the errors in
%this measurement.
%
%
  Preliminary polarization
measurements~\cite{Castromonte} with this data indicate a much smaller polarization.
Therefore  the effect on our measurement of a possible
\lambdacm polarization is neglible.

The fitting
conditions and the cuts were changed in a reasonable manner,
on the whole data set.
The \alm and \A values obtained by these variants
are all {\em a priori} equally likely, therefore this 
uncertainty can be
estimated by the {\it r.m.s.} of the measurements~\cite{brkkpipi}.

\begin{figure}[!!t]
\begin{center}
\epsfysize=6.5cm
\epsfxsize=6.5cm
\epsfbox{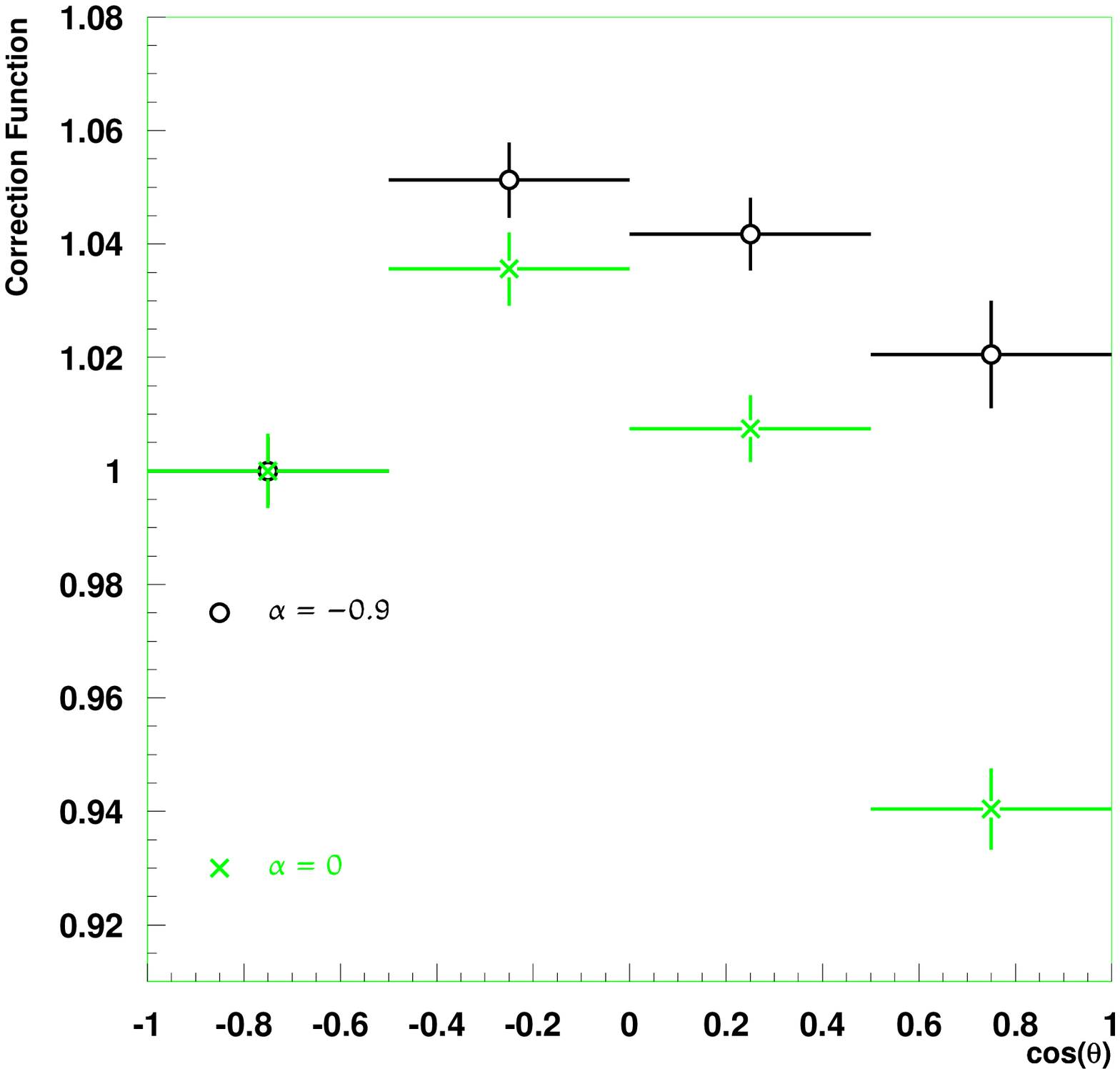}
%\vspace{0.5cm}
\end{center}
\caption{
The correction function $\epsilon(\theta_i,\alpha)$
corresponding to $\al\times\alpha_\Lambda=-0.9$ (circles)
and to $\al\times\alpha_\Lambda=0$ (crosses)
for the \lambdacm sample,
plotted superimposed to each other. In this figure
$\alpha $ means actually $\al\times\alpha_\Lambda$.
All the correction functions
corresponding to intermediate values, for any given
\costm interval,
lie between the values plotted on this figure.
}
\label{corrcompn}
\end{figure}

\begin{figure}[!!t]
\begin{center}
\epsfysize=6.5cm
\epsfxsize=6.5cm
\epsfbox{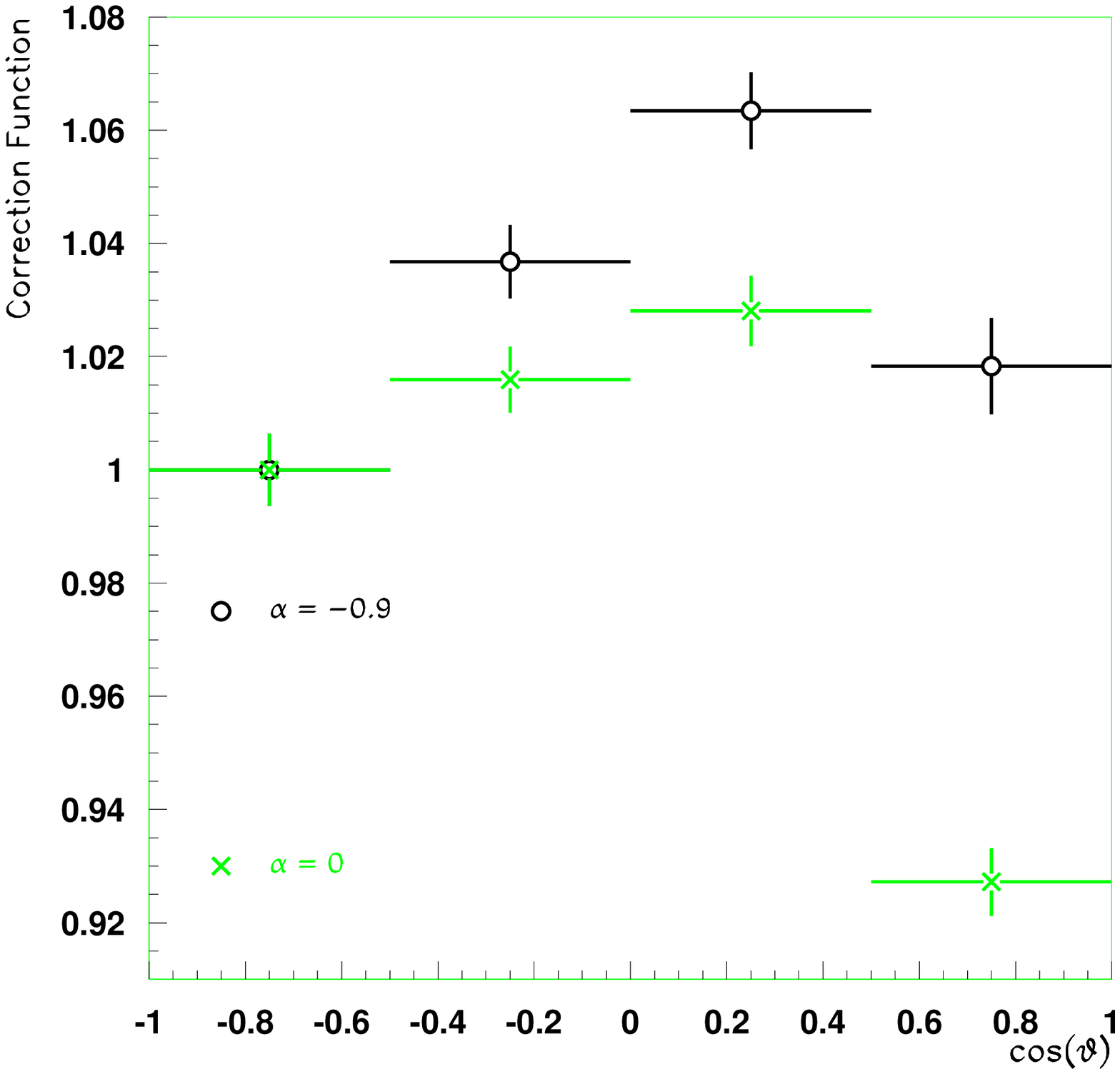}
%\vspace{0.5cm}
\end{center}
\caption{
The correction function $\epsilon(\theta_i,\alpha)$
corresponding to $\alb\times\alpha_{\overline \Lambda}=-0.9$ (circles)
and to $\alb\times\alpha_{\overline \Lambda}=0$ (crosses)
for the \lambdacbm sample,
plotted superimposed to each other. In this figure
$\alpha $ means actually $\alb\times\alpha_{\overline \Lambda}$.
All the correction functions
corresponding to intermediate values, for any given
\costm interval,
lie between the values plotted on this figure.
}
\label{corrcompnbar}
\end{figure}

\begin{figure}[!!t]
\begin{center}
\epsfysize=8.cm
\epsfxsize=8.cm
\epsfbox{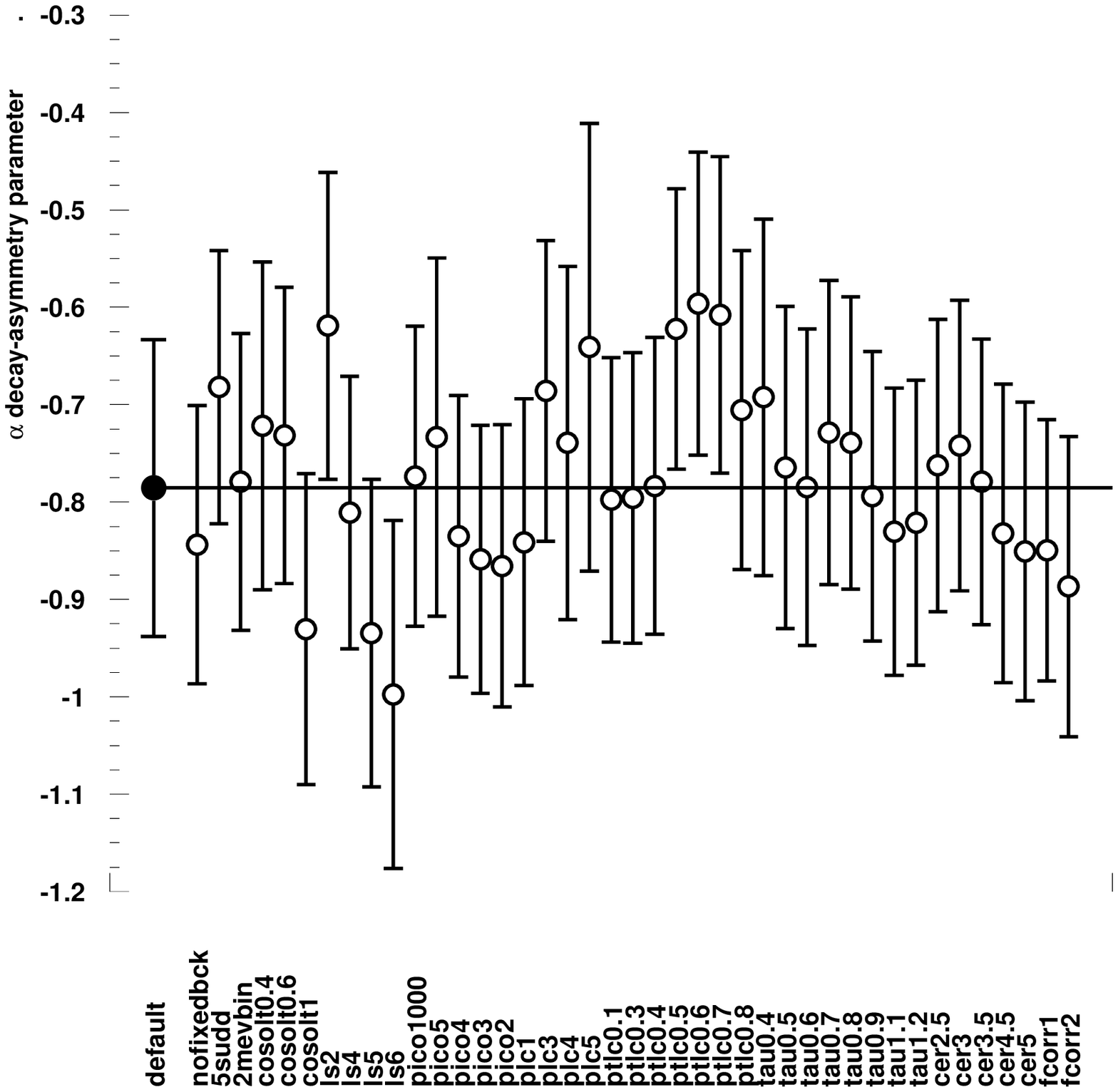}
%\vspace{0.5cm}
\end{center}
\caption{
Summary of the systematic checks
performed on $\al$, calculated in the hypothesis
of \emph{CP} conservation. The first circle on the left of
plot is the result quoted in this paper.
}
\label{systalfa}
\end{figure}

\begin{figure}[!!t]
\begin{center}
\epsfysize=8.cm
\epsfxsize=8.cm
\epsfbox{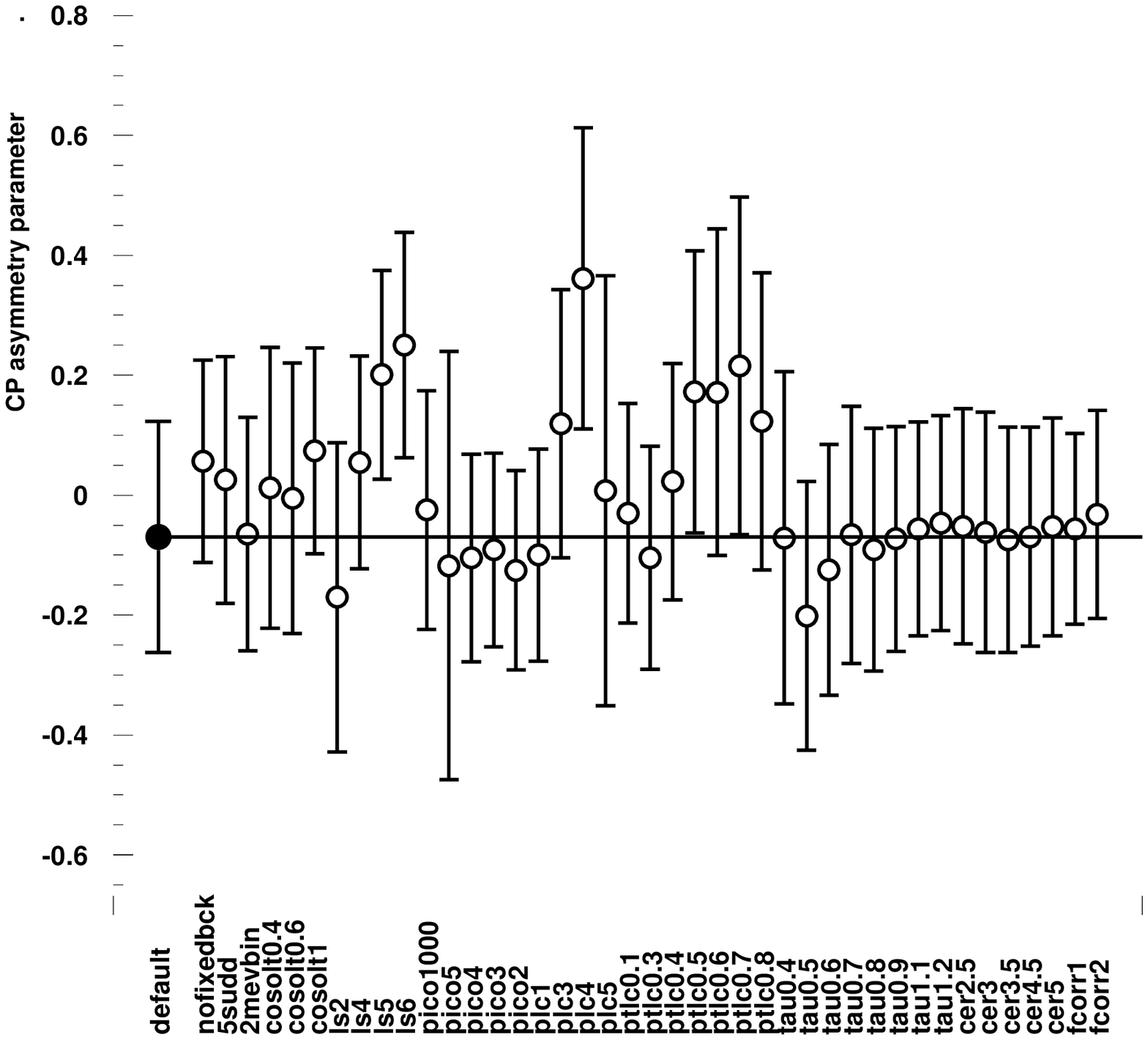}
%\vspace{0.5cm}
\end{center}
\caption{
Summary of the systematic checks
performed on $\mathcal{A}$, calculated in the hypothesis
of \emph{CP} conservation. The first circle on the left of
plot is the result quoted in this paper.
}
\label{systA}
\end{figure}

As explained earlier,
the background shape of each $\Lambda\pi^+$ mass plot was fit with
a second degree polynomial extracted from the
wrong sign sample and only an overall scale factor was varied in the mass fit.
To check how critical this assumption is,
a mass fit was performed in which all twelve polynomial coefficients
(three coefficients times four \costm bins)
were fit free parameters instead of taking the shape from
the wrong sign sample.
The results of the fit performed in this way are
$\mathcal{A} = \Avalnofix\pm \erAvalnofix $ and
$\al = \valalnofix \pm \ervalalnofix$ when one assumes \emph{CP} invariance.
These results are in agreement with
those of the standard analysis
and show that the result is not sensitive to the way the
background is fit.
The following further systematic checks were performed:
\begin{itemize}
\item
using a 2 $\mathrm{MeV}/c^2$ binning in the mass plots instead
 of the 5 $\mathrm{MeV}/c^2$
 binning
of the standard analysis;
\item
dividing the \costm range in 5 equally wide intervals instead of 4 intervals;
\item
varying the most relevant selection cuts in order to check how well
the Monte Carlo reproduces the experimental apparatus, namely~:
variation of the $l/\sigma$ cut;
variation of the cut on the transverse (with respect to the
beam direction) momentum of the $\lambdac$;
 variation of the cut on $|\cos\omega|$;
sharpening of the {\it picon} cut;
 variation of the cut on the proper time of the \lambdacm candidate;
variation of the \v{C}erenkov cut on the proton.
\item
Another check was performed to assess the possible systematic error
due to errors in the correcting function
%due to the imprecision of the knowledge of the correcting function
$\epsilon(\theta_i,\alpha)$
caused both by the interpolation method and by an intrinsic
failure of the Monte Carlo to simulate the experimental
apparatus. The correction functions that differ the most are those
corresponding to the extreme values of $\al\, \alpha_{\Lambda}$,
(or $\alb\, \alpha_{\overline \Lambda}$) namely
$\al\, \alpha_{\Lambda}=-0.9$ and $\al\, \alpha_{\Lambda}=0$
(and correspondingly for the \lambdacbm sample)
as shown by Figs.~\ref{corrcompn} and \ref{corrcompnbar}.
The fit was performed once imposing
$\al\, \alpha_{\Lambda}=\alb\, \alpha_{\overline \Lambda}=-0.9$
(no interpolation) and then imposing
$\al\, \alpha_{\Lambda}=
\alb\, \alpha_{\overline \Lambda}=0$.
These results were taken as an indication of the magnitude
of a possible systematic error.
\end{itemize}
\noindent
The results for \alm and \A for all these systematic
checks are summarized in Figs.~\ref{systalfa} and \ref{systA}. As 
explained
previously the $r.m.s.$ of all those results is taken as the estimate of the
systematic error. For \alm this is $\pm 0.12$ and for \A it is $\pm 0.13$
%%%%%%%%%%%%%%%%% minimontecarlo %%%%%%%%%%%%%%%%%%%%%%%%%%%%%%%
\subsection{Check of the statistical
errors returned by the fitter and possible bias}
A check was performed to assure that the errors returned by the fit were
realistic and to assess if there was a bias in the fitting procedure. For both the
particle and the antiparticle cases, the  bin
contents of the four \costm region mass plots (Fig.~\ref{cost_n} and
\ref{cost_nbar})  and the four
wrong sign plots
% (Fig.~\ref{wrongsign} and \ref{wrongsignbar})
were randomly varied according to Poisson statistics 1000 times.
%%%%%%%%%%%%%%%%%%%%%%%%%%%%%%%%%%%%%%%
Each set of eight "varied" plots was analyzed with the fitting program as it
were regular data and the values of \A and \alm extracted.
With this method we  assessed a bias of $0.000 \pm 0.005$ on 
\A and we simply added $0.005$ in quadrature to the systematic error.
For \alm we assessed a bias of $-0.010\pm 0.005$ and a slight underestimation of the
statistical error returned by the fit (0.16 instead of 0.15). Consequently this bias
 was
subtracted, an error of $0.005$ was added in quadrature to the systematic error and
the statistical error was increased to $0.16$.
%%%%%%%%%%%%%%%%%%%%%%%%%%%%%%%%%%%%%%%%%
%
% These 1000 values are
%plotted in Fig.~\ref{Aminimc} and \ref{alminimc} respectively.
%Superimposed in these plots are the Gaussians fitting them.
% If the fit was
%unbiassed, we would expect the mean value of each Gaussian to correspond exactly
%to the values returned by the original fit. Actually that is the case for \A
% (the mean of the Gaussian of Fig.~\ref{Aminimc} is $-0.07\pm 0.005$, the same
%value of the original fit). The uncertainty of $0.005$ is added in quadrature to the systematic
%error.
%For \alm the mean of the Gaussian is
%$-0.80$ when the original fit value is $\valal$ showing a bias of $\albias \pm 0.005$.
%This bias is only marginally different from zero, but
%will be taken out from the fit original value (which then becomes
% $\al = \valalbis$). The uncertainty of $0.005$ on the bias is included in the total systematic
% error.
%
%Moreover, if the statistical errors on \A and \alm
%returned by the fit were correct, they would
%coincide with the $\sigma$ of the corresponding Gaussian.
%That is the case for \A (both the $\sigma$ of the Gaussian and the original fit
%error are $\erstAvalbis$) while for \alm the fit slightly underestimates the error
%(the $\sigma$ of the Gaussian is $\erstvalalbis$ while the original fit
%error is $\erstvalal$). Consequently the statistical error
% on \alm is increased to $\erstvalalbis$.
\section{Final results and
comparison with previous results and with the theoretical predictions}
After all systematic checks,  bias correction and error inflation
the final results of the measurements are
$\al = \valalbis \pm \erstvalalbis \pm \ersyvalal$ and
$\mathcal{A} = \Avalbis\pm \erstAvalbis \pm \ersyAval$.
Previous results obtained for \alm are listed in Table~\ref{tabone}.
\begin{table}
\begin{center}
\begin{tabular}{|l|c|}
\hline
Experiment& \alm \\ \hline
FOCUS (this result)& $\valalbis \pm \erstvalalbis \pm \ersyvalal  $     \\
CLEO~II~\cite{CLEO2}
   & $-0.94  \pm 0.21  \pm 0.12  $    \\
ARGUS~\cite{ARGUS} & $-0.96   \pm 0.42   $      \\
CLEO~\cite{CLEO}   & $-1.1  \pm 0.4    $      \\ \hline
\end{tabular}
\caption{Comparison of the \alm result with other experiments.}
\label{tabone}
\end{center}
\end{table}
The present analysis result is a higher precision
and it is in agreement with earlier results.
No previous measurements for \A have been reported.

\noindent
Theoretical predictions of the two-body baryon-pseudoscalar decays
 of charmed baryons have
been made by a number of authors in the past.
\begin{table}
\begin{center}
\begin{tabular}{|c|c|}
\hline
 Paper & \alm \\ \hline
Ivanov {\it et al.}~\cite{ivanov}(1998) & $-0.95 $     \\
Zenczykowski~\cite{zen1}(1994)& $-0.99$    \\
Zenczykowski~\cite{zen2}(1994)& $-0.86$    \\
Uppal {\it et al.}~\cite{uppal}(1994) & $-0.85 $     \\
Cheng and Tseng~\cite{cheng1}(1993) & $-0.95 $     \\
Cheng and Tseng~\cite{cheng2}(1992) & $-0.96 $     \\
Xu and Kamal~\cite{xu}(1992) & $-0.67 $     \\
K\"orner and Kramer~\cite{korner}(1992) & $-0.70 $     \\
 \hline
\end{tabular}
\caption{
Summary of the most recent theoretical predictions on \alm (under
the \emph{CP} conservation hypothesis)
}
\label{predizioni}
\end{center}
\end{table}
They predict the
parity conserving amplitude
$ A_+$ and parity violating amplitude $A_-$  of Eqn. (\ref{cp_viola})
from which \alm  and the decay rate can be derived.
The most recent predictions are listed in Table~\ref{predizioni}. All
predictions are in agreement within errors with the FOCUS
experimental result.
\section{Conclusions}
 Using data from the FOCUS (E831) experiment at Fermilab,
we have studied the
decay mode $\signal$ and measured the asymmetry parameter
\alm (under the \emph{CP} conservation hypothesis) and the \emph{CP} parameter
\A. Our result for \alm is consistent with previous measurements and is
more accurate.  Our result for \A is the first measurement of this
quantity
and is consistent with
zero.
%%%%%%%%%%%%%%%%%%%%%%%%%%%%%%%%%%%%%%%%%%%%%%%%%%%%%%%%%%%%%%%%%%%%%%%%%%%%%%%
%
%                                  acknowledgements
%
%%%%%%%%%%%%%%%%%%%%%%%%%%%%%%%%%%%%%%%%%%%%%%%%%%%%%%%%%%%%%%%%%%%%%%%%%%%%%%%%

\vspace{1.cm}

We wish to acknowledge the assistance of the staffs of Fermi National
Accelerator Laboratory, the INFN of Italy, and the physics departments of
the collaborating institutions. This research was supported in part by the
U.~S. National Science Foundation, the U.~S. Department of Energy, the
Italian Istituto Nazionale di Fisica Nucleare and Ministero della Istruzione
Universit\`a e Ricerca, the Brazilian Conselho Nacional de Desenvolvimento
Cient\'{\i}fico e Tecnol\'ogico, CONACyT-M\'exico, and the Korea Research
Foundation of the Korean Ministry of Education.

%%%%%%%%%%%%%%%%%%%%%%%%%%%%%%%%%%%%%%%%%%%%%%%%%%%%%%%%%%%%%%%%%%%%%%%%%%%%%%%
%
%                                  REFERENCES
%
%%%%%%%%%%%%%%%%%%%%%%%%%%%%%%%%%%%%%%%%%%%%%%%%%%%%%%%%%%%%%%%%%%%%%%%%%%%%%%%

%%%%%%%%%%%%%%%%%%%%%%%%%%%%%%%%%%%%%%%%%%%%%%%%%%%%%%%%%%%%%%%%%%%%%%%%%%%%%%%
%
%                                  FIGURE
%
%%%%%%%%%%%%%%%%%%%%%%%%%%%%%%%%%%%%%%%%%%%%%%%%%%%%%%%%%%%%%%%%%%%%%%%%%%%%%%%

\newpage

%\begin{figure}[!!t]
%\begin{center}
%\epsfysize=5.0cm
%\epsfxsize=5.5cm
%\epsfbox{massfig.eps}
%\end{center}
%\caption{$\Lambda\pi$ effective mass after the selection.
%a) total sample; b) $\Lambda\pi^+$ sample; c) $\overline{\Lambda}\pi^-$ sample.}
%\label{mass}
%\end{figure}
\newsavebox{\plotone}
\savebox{\plotone}{\epsfig{file=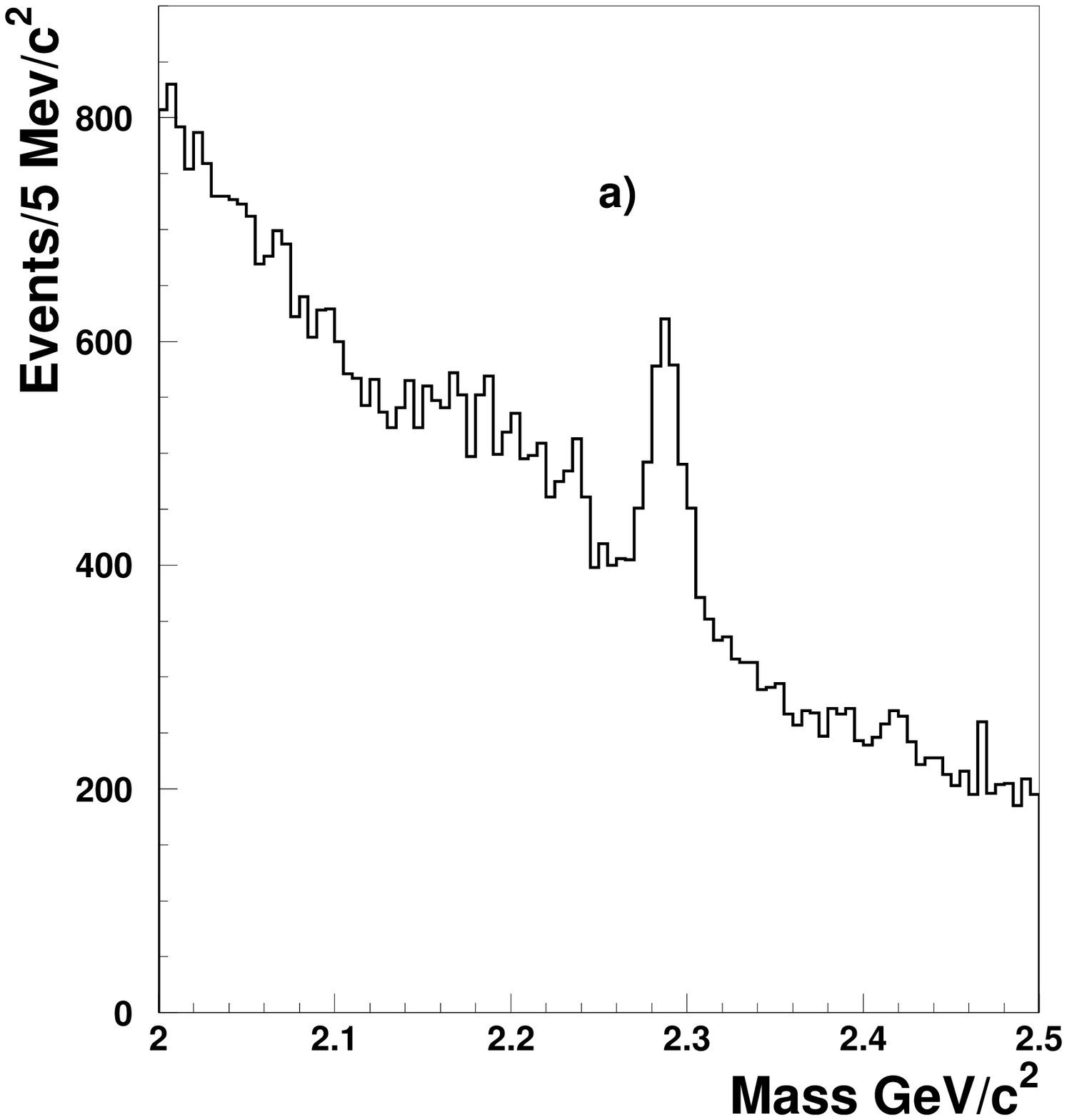,width=7cm}}
\newsavebox{\plotdue}
\savebox{\plotdue}{\epsfig{file=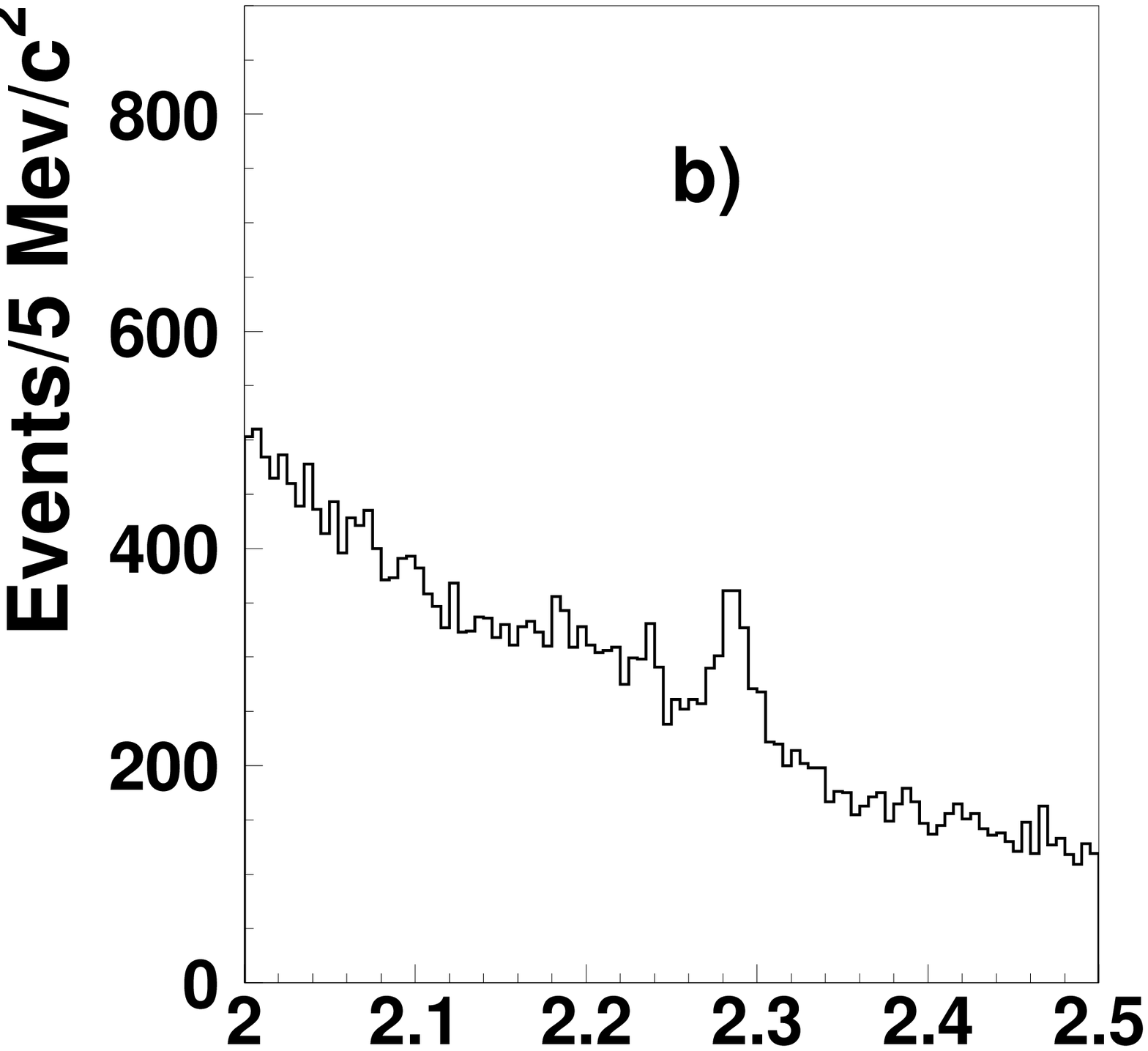,width=3.3cm}}
\newsavebox{\plottre}
\savebox{\plottre}{\epsfig{file=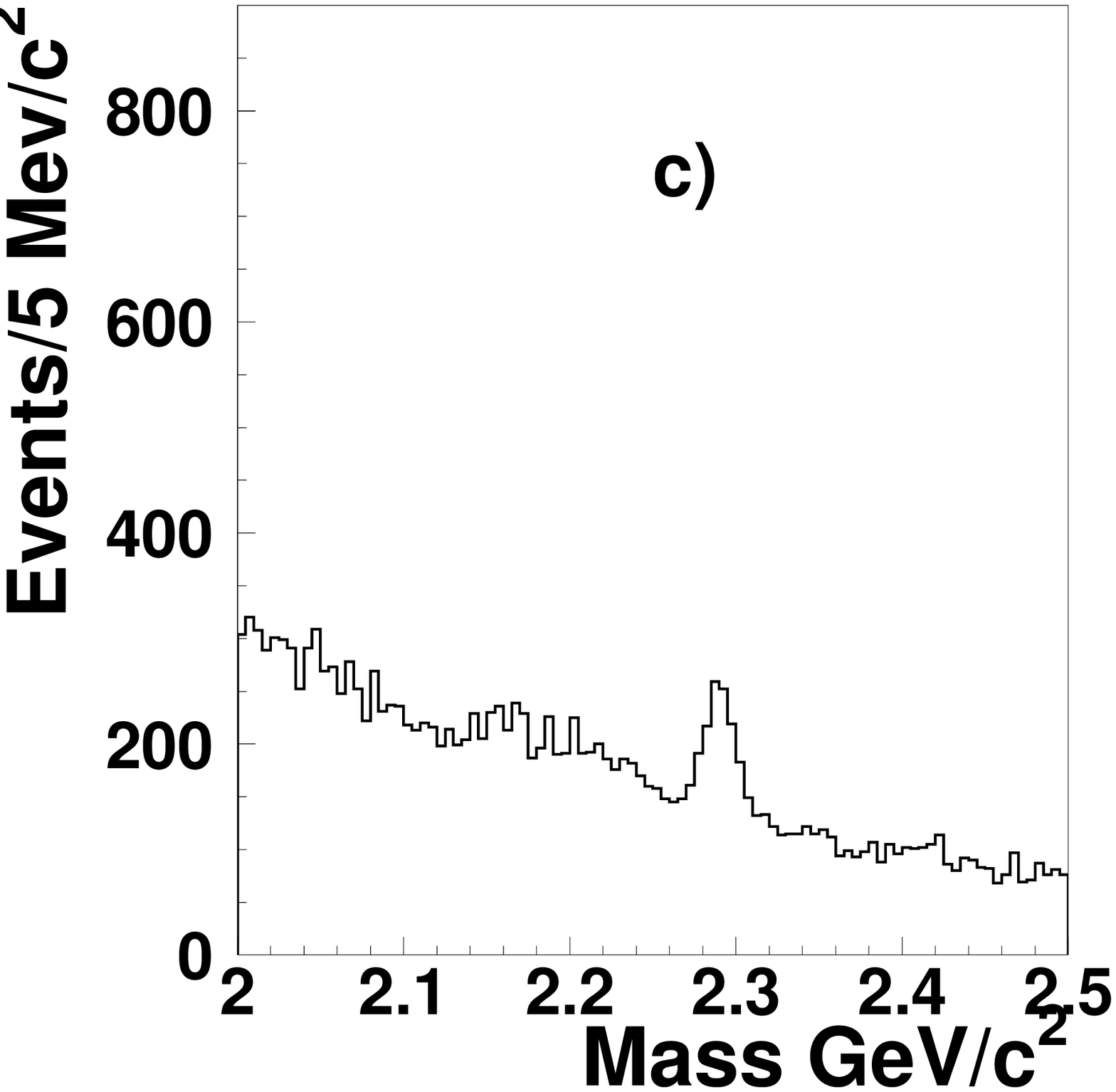,width=3.3cm}}

\begin{figure}[!!t]
\begin{picture}(400,150,)(0,0)
 \put(40,0){\usebox{\plotone}}
 \put(240,95){\usebox{\plotdue}}
 \put(240,10){\usebox{\plottre}}
%\put(0,0){a}
%\put(400,0){b}
%\put(0,150){c}
%\put(400,150){d}
\end{picture}
\caption{$\Lambda\pi$ effective mass after the selection.
a) total sample; b) $\Lambda\pi^+$ sample; c) $\overline{\Lambda}\pi^-$ sample.}
\label{mass}
\end{figure}
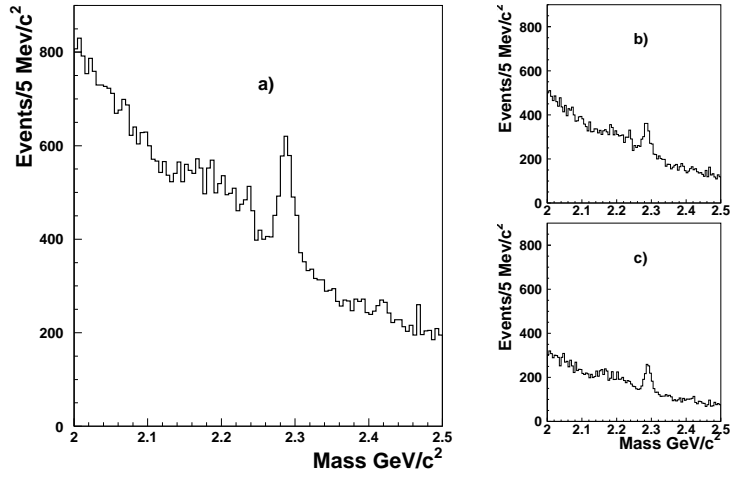

\begin{figure}[!!t]
\begin{center}
\epsfysize=6.5cm
\epsfxsize=6.5cm
\epsfbox{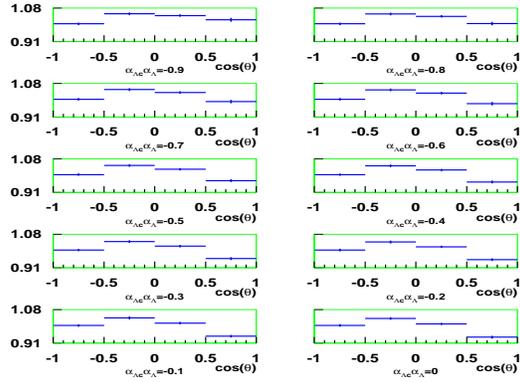}
%\vspace{0.5cm}
\end{center}
\caption{Correction function $\epsilon(\theta_i,\alpha)$
calculated from Monte Carlo,
for the \lambdacm sample,
at ten values of $\al\times\alpha_\Lambda$, spanning
the range from $-0.9$ to 0. All correction functions
are normalized such that they are 1 in the first \costm
bin. Note that an
overall normalizing factor can be chosen for the correction functions
without changing the result of the fit.
}
\label{corrfuncn}
\end{figure}

\begin{figure}[!!t]
\begin{center}
\epsfysize=6.5cm
\epsfxsize=6.5cm
\epsfbox{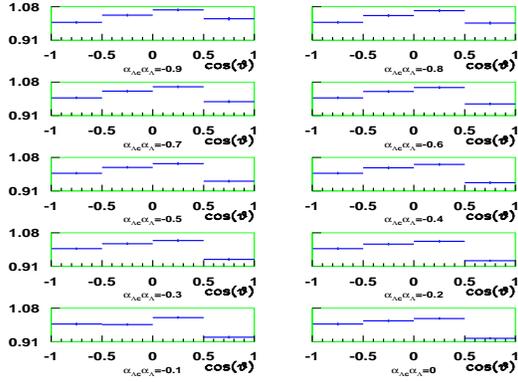}
%\vspace{0.5cm}
\end{center}
\caption{Correction function $\epsilon(\theta_i,\alpha)$
calculated from Monte Carlo,
for the \lambdacbm sample,
at ten values of $\alb\times\alpha_{{\overline\Lambda}}$, spanning
the range from $-0.9$ to 0. All correction functions
are normalized such that they are 1 in the first \costm
bin.
}
\label{corrfuncnbar}
\end{figure}

%\begin{figure}[!!t]
%\begin{center}
%\epsfysize=6.5cm
%\epsfxsize=6.5cm
%\epsfbox{minimc/asy.ps}
%\end{center}
%\caption{
%The thousand values of A obtained varying in Poisson way the data mass plots are
%plotted, with superimposed a Gaussian fit. The Gaussian fit has a mean of
%$-0.07$ and $\sigma = 0.19$ to be compared with the result of the data fit that
%gives $\mathcal{A}=-0.07\pm 0.19$, thereby showing there is no bias in the fit
%procedure and the statistical errors are correct.
%}
%\label{Aminimc}
%\end{figure}

%\begin{figure}[!!t]
%\begin{center}
%\epsfysize=6.5cm
%\epsfxsize=6.5cm
%\epsfbox{minimc/alfa.ps}
%\end{center}
%\caption{
%The thousand values of \alm
% obtained varying in Poisson way the data mass plots are
%plotted, with superimposed a Gaussian fit. The Gaussian fit has a mean of
%$-0.80$ and $\sigma = 0.16$ to be compared with the result of the data fit that
%gives $\mathcal{A}=-0.79\pm 0.15$, thereby showing there is a bias
%of 0.01  in the fit
%procedure and the statistical errors are slightly underestimated.
%}
%\label{alminimc}
%\end{figure}

\end{document}